# Are Social Networks Really Balanced?

Ernesto Estrada[1,2,3] and Michele Benzi[4]

[1]Department of Mathematics & Statistics, [2]Institute of Complex Systems at Strathclyde, University of Strathclyde, Glasgow G1 1XH, UK, [3]Institute for Quantitative Theory and Methods (QuanTM), Emory University, [4]Department of Mathematics and Computer Sciences, Emory University, Atlanta, GA 30322, USA.

**There is a long-standing belief that in social networks with simultaneous friendly/hostile interactions (signed networks) there is a general tendency to a global balance. Balance represents a state of the network with lack of contentious situations. Here we introduce a method to quantify the degree of balance of any signed (social) network. It accounts for the contribution of all signed cycles in the network and gives, in agreement with empirical evidences, more weight to the shorter than to the longer cycles. We found that, contrary to what is believed, many signed social networks—in particular very large directed online social networks—are in general very poorly balanced. We also show that unbalanced states can be changed by tuning the weights of the social interactions among the agents in the network.**

Social networks represent a large proportion of the complex socio-economic organization of modern society. They represent social entities, such as countries, corporations or people, interconnected through a wide range of social ties, which include political treaties, commercial trade, friendship and collaboration, among others.[1,2] In recent years a new dimension of social networks has emerged with the development of online social communities, which contribute and share contents on the WWW.[3-6] In many of these scenarios the interactions among the social entities go beyond the simple connected-disconnected networks, e.g., friend-not friend relationship, to include antagonistic relations among the connected entities. These are the cases in which social entities can display, for instance, ally/enemy, friend/foe, trust/distrust relationships. In these cases the social system must be represented as a signed network in which the edges of the network can be either positive (+) to denote ally, friendship, trust, or negative (−) to denote enemy, foe, distrust.[7-14]

The origin of the study of signed networks can be traced back to the work of Heider,[15] who formulated a theory of social balance to understand the causes of tensions and conflicts in networks where friendship/animosity relations coexist. The use of signed networks was then proposed by Cartwright and Harary[16] to model the existence of balance/unbalance in such social systems. The lack of balance in a signed network is produced by the existence of groups of individuals cyclically connected where the number of negative edges is odd.[16-20] For instance, a triad in which Bob and Sue are friends with Mike but are unfriendly with each other is believed to be destabilized by the attempts of Bob (Sue) to strengthen his (her) relation with Mike by suggesting he (she) breaks with Sue (Bob). This unpleasant situation is believed to catalyze a change in the social relations to produce a balanced state in the network.

A signed network is balanced if and only if all its cycles are positive, where the sign of a cycle is the product of the signs of its edges. This black-and-white consideration of network balance has been widely studied and documented in social systems for many years. Only recently, gray-scales in which the quantitative determination of how unbalanced a social network is, have been considered in the literature.[8-10] Some of these approaches consider only triads to account for balance, which excludes the contribution to unbalance of longer cycles,[8] or do not provide local information about individual contribution to balance.[9] A method for computing the degree of unbalance of a signed network was proposed by Facchetti et al.[10] by using ground-state calculations in large-scale Ising spin glasses. Using their approach for undirected versions of three online social networks, they have concluded that "*currently available networks are indeed extremely balanced*".[10] This conclusion agrees very well with Heider balance theory. In previous work, Leskovec et al.[8] have analyzed the statistical significance of all possible triads in the same online social networks. Their results contrast very much with those of Facchetti et al.[10] as they found that the abundance of certain signed triads does not follow Heider's theory and is more in line with Davis's weaker notion of balance,[21] which states that only the triangles with two positive edges ("*the enemy of my enemy is my friend*") are implausible in real social networks, but all other triangles are permissible. When the more realistic directed versions of these networks were considered by Leskovec et al.[8] they concluded that "*many of basic predictions of balance theory no longer apply*". On the one hand, Leskovec et al.[8] considered only signed triads, which are arguably the most important fragments in determining balance but not the only ones. On the other hand, the method of Facchetti et al.[10] gives the same importance to the lack of balance in every cycle, indistinctly of its length. This contrasts with the well-documented fact that the longer cycles have less effect upon a person's tension than the shorter ones.[22]

These discrepancies are not only on the quantitative side of the problem but also in the conceptual one. Using the previous hypothetical example, it is plausible that Mike feels comfortable by acting as a mediator in the disputes between Bob and Sue, and that Bob (Sue) feels certain stability in the use of Mike to influence Sue's (Bob's) opinions in her (his) favor. This situation was indeed considered by Heider[23] already in 1958 when he wrote that "*there may also be a tendency to leave the comfortable equilibrium, to seek the new and adventurous. The tension produced by unbalanced situations often has a pleasing effect on our thinking and aesthetic feelings. Balanced situations can have a boring obviousness and a finality of superficial self-evidence. Un-balanced situations stimulate us to further thinking; they have the character of interesting puzzles, problems which make us suspect a depth of interesting background*." Then, the correct determination of the degree of balance of real-world signed social networks is of vital importance to empirically validate one of these hypotheses over the other.

Here we consider a new way to quantify the degree of balance in a signed network, which accounts for the contribution of all signed cycles in the network, by giving more weights to the shorter than to the longer ones. This method can be formulated as an equilibrium constant for a hypothetical equilibrium between the real-world signed network and its underlying unsigned version. Using this approach we study five signed social networks of different sizes and representing very different social scenarios. We found that many of these networks, in particular the large online social networks, are very far from balance. Furthermore, we also show that the level of balance that a network displays can be significantly changed by tuning the weights of the social links among the connected actors in the network. The approach developed in this work is easy to implement computationally even for very large networks, as in general its complexity scales linearly with the size of the network.

## Results

**Walk balance for signed networks.** We consider here directed (undirected) signed networks $\Sigma=(V,E)$ in which the weight of every edge is $+1$ or $-1$. Every signed directed (undirected) network has an underlying unsigned network, which consists of the same set of nodes and edges as $\Sigma$ with all edges having positive sign. The underlying network of $\Sigma$ is represented here by $|\Sigma|$.[24] In this work we denote by $n$ the number of nodes and by $m$ ($m^+, m^-$) the number of (positive, negative) edges. Let $A(\Sigma)$ and $A(|\Sigma|)$ be the adjacency matrices of the signed (directed) network and its underlying unsigned graph, respectively. A directed (undirected) walk of length $k$ in $\Sigma$ is a sequence of (not necessarily distinct) nodes $v_0, v_1, \cdots, v_{k-1}, v_k$ such that for each $i = 1,2,\ldots,k$ there is a link from $v_{i-1}$ to $v_i$. If $v_0 = v_k$, the walk is called a *closed walk*. The sign of a walk is the product of the signs of all the edges involved in it.[24] We remind that in a (directed) network, the total number of walks of length $k$ is given by $tr(A^k)$, where $A$ is the adjacency matrix of the graph and $tr$ is the trace of the matrix. A *balanced weighted closed walk* (BCW) is a closed walk of length larger than zero with a positive sign. Similarly, an *unbalanced weighted closed walk* (UCW) is a closed walk of length larger than zero with negative sign.

We recall that a signed directed network is called (cycle) balanced if every cycle of it is positive. We introduce now the following definition: A signed (directed) network is said to be walk-balanced if every (directed) closed walk of it is positive. Obviously, a cycle balanced network is also a walk-balanced one and *vice versa*. The main difference arises in the quantification of how close to balanced is an unbalanced network. We start by considering that in social networks it has been empirically demonstrated that the longer cycles have less effect upon a person's tension than the shorter ones. Then, we introduce here a weighted sum of all closed walks in a directed signed network which takes into account this empirical observation. That is, we consider $D(\Sigma) = \sum_{k=0}^{\infty} tr\left[A(\Sigma)^k\right]/k!$, which converges to $D(\Sigma) = tr\exp(A(\Sigma))$. Due to the fact that every BCW contributes positively to $D(\Sigma)$ and that every UCW contributes negatively, we have that $tr\left(e^{A(\Sigma)}\right) = \mu^B - |\mu^U|$, where $\mu^B$ ($\mu^U$) is the sum of the weighted (by inverse factorial of the length) balanced (unbalanced) closed walks, and $|\cdots|$ represents the absolute value. Similarly, we can consider the same term in the underlying graph $|\Sigma|$, which results in $tre^{A(|\Sigma|)} = \mu^B + |\mu^U|$. Next, we define

$$K = \frac{tr\exp\left(A(\Sigma)\right)}{tr\exp\left(A(|\Sigma|)\right)} = \frac{\sum_{j=1}^{n}\exp\left(\lambda_j(\Sigma)\right)}{\sum_{j=1}^{n}\exp\left(\lambda_j(|\Sigma|)\right)}, \tag{1}$$

where $\lambda_j(\Sigma)$ and $\lambda_j\left(|\Sigma|\right)$ are the eigenvalues of $A(\Sigma)$ and $A(|\Sigma|)$, respectively. It is straightforward to realize that

$$K = \frac{\mu^B - |\mu^U|}{\mu^B + |\mu^U|}. \tag{2}$$

This means that the ratio of unbalanced to balanced CWs can be obtained as,

$$\mathscr{U} = \frac{|\mu^U|}{\mu^B} = \frac{1-K}{1+K}, \tag{3}$$

which represents the extent of the lack of balance in a given signed network. For instance, a network is highly unbalanced if $K \approx 0$, which makes $\mathscr{U} \approx 1$. On the other hand, a balanced network has $\mathscr{U} = 0$.

It has been proved that $\lambda_j(\Sigma) = \lambda_j(|\Sigma|)$ (with multiplicities) if and only if the signed network $\Sigma$ is balanced.[25] Consequently, $K \leq 1$, with equality if and only if the signed network is balanced. As the network departs from balance the walk-balance index drops down to 0. That is, $K$ tends asymptotically to zero for certain classes of graphs which will be called *maximally unbalanced networks* (see SI). Thus, $0 < K \leq 1$, with values close to unity indicating more balance in the network, values close to 0 for largely unbalanced networks.

In an analogous way as for the definition of spectral balance for the whole network we define the following index that characterizes the degree of balance of a given node:

$$K_i = \left(\left(\exp\left(A(\Sigma)\right)\right)_{ii} - 1\right) / \left(\left(\exp\left(A\left(|\Sigma|\right)\right)\right)_{ii} - 1\right). \tag{4}$$

For the sake of comparison we will use here the ratio of the number of signed to unsigned triangles: $K_3 = tr\left(A(\Sigma)^3\right)/tr\left(A(|\Sigma|)^3\right)$, which can be written as $K_3 = (t_B - t_U)/(t_B + t_U)$, where $t_B$ and $t_U$, are the number of balanced and unbalanced triangles, respectively.

**Global balance as an equilibrium constant.** We consider here a hypothetical dynamical system in which an unsigned network $|\Sigma|$ changes the sign of a few links to give rise to a signed network $\Sigma$ (see Fig. 1). This is the network analogous of a conformational change in a molecule, such as the conformational change in a protein or DNA. To complete this analogy we need to assume that the network is submerged into a thermal bath with inverse temperature $\beta = (k_B T)^{-1}$, where $k_B$ is the Boltzmann constant.[26]

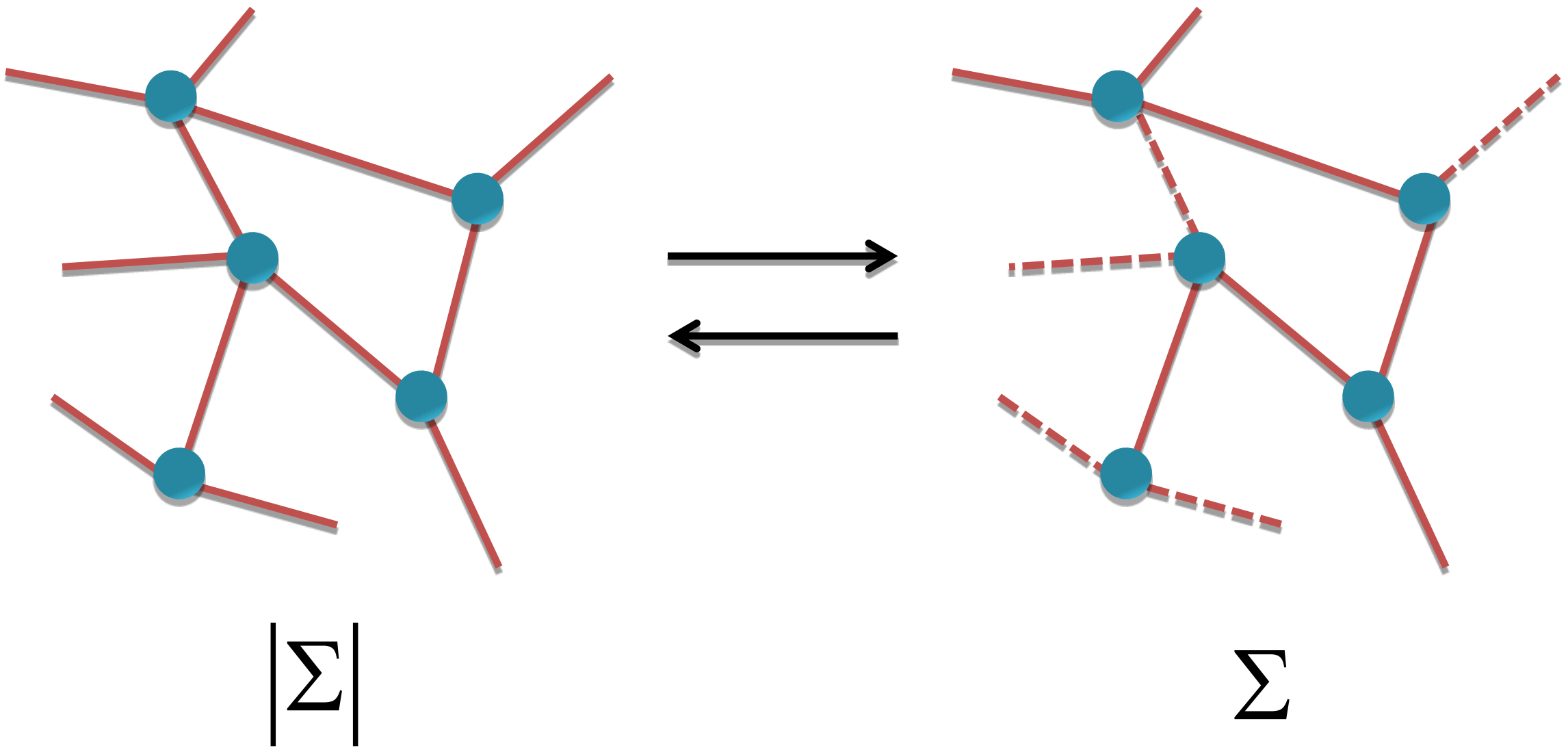

**Fig. 1.** Hypothetical equilibrium between a signed graph $\Sigma$ and its underlying unsigned graph $|\Sigma|$. Continuous lines represent positive and dashed lines represent negative links.

The change of the free energy of the thermodynamic process is the difference of free energies of the final and initial states: $\Delta F = F_\Sigma - F_{|\Sigma|}$, where $F_\Sigma = -\beta^{-1} \ln Z_\Sigma$, $F_{|\Sigma|} = -\beta^{-1} \ln Z_{|\Sigma|}$. The corresponding partition functions are $Z_\Sigma = tr\left(e^{\beta A(\Sigma)}\right)$, and $Z_{|\Sigma|} = tr\left(e^{\beta A(|\Sigma|)}\right)$, in which $A(\Sigma)$ and $A(|\Sigma|)$ are the adjacency matrices of the signed and unsigned graphs, respectively. It is straightforward to realize that the change in free energy of the system is given by $\Delta F = -\beta^{-1} \ln\left(Z_\Sigma / Z_{|\Sigma|}\right)$, and we can write the equilibrium constant for the process represented in Fig. 1 as the ratio of the two partition functions $K = \exp\left(-\beta \cdot \Delta F\right) = Z_{\Sigma} / Z_{|\Sigma|}$. Consequently, the equilibrium constant is written as

$$K\left(\beta\right) = \frac{tr\left(\exp\left(\beta A(\Sigma)\right)\right)}{tr\left(\exp\left(\beta A\left(|\Sigma|\right)\right)\right)}, \tag{5}$$

which means that the walk-balance index is just a particular case of this equilibrium constant for $\beta = 1$.

**Balance in small social networks.** We start the analysis of some real-world signed networks by considering two systems formed by small networks. The first deals with the evolution of the relations among the major players in the World War I (WWI).[7] The second is provided by the Gahuku-Gama subtribe system of the Eastern Central Highlands of New Guinea.[27,28] In Fig. 2 we represent the six protagonists of WWI at different time snapshots, starting from the Three Emperors' league in 1872 and ending with the British-Russian Alliance of 1907. As can be seen the general trend is towards increasing the balance in time. In 1872 the global balance index is $K = 0.4668$, which is increased up to $K = 0.5489$ in 1904, just before a total balance is produced in 1907 with the British-Russian Alliance. This trend is broken with the break of the Russia-Germany alliance in 1890 which makes the global balance drops to $K = 0.4681$. If instead of $K$ we consider only the contribution of triads to the global balance, i.e., by means of $K_3$, it looks like the German-Russian Lapse caused a dramatic decrease in the global balance. That is, the values of $K_3$ for the six signed networks are: 0.428, 0.500, 0.200, 0.500, 0.500, 0.500, 1.000. What happens is that although a large triad unbalance exists in this period, it is compensated somehow in some tetrad balance. For instance, AH-It-Ge-Fr-AH, Ru-AH-It-Ge-Ru, and Fr-GB-Ru-AH-Fr are examples of balance squares which compensate the lack of triad balance. Consequently, the consideration of all cycles, like in the walk-balance approach, is more appropriate than the triads-only methods for having a correct picture of global balance in social networks. We also calculate the local contribution of each country to the balance. As can be seen in Fig. 2, Germany (Ge), the Austro-Hungarian (AH) empire and Italy (It) always display a large balance across time, while Great Britain (GB), France (Fr) and Russia (Ru) where always more unbalanced. It should be noticed that after 1882 the alliance between Ge, AH and It was permanent while the three other major players (GB, Fr and Ru) where changing their alliances and enmities all the time, until the formation of the British-Russian alliance of 1907. After this point, when all the countries were balanced, the WWI started, maybe as a consequence of the fact that every country felt strong enough to go to war. As remarked by Antal et al. "*while social balance is a natural outcome, it is not necessarily a good one!*".[7] Although these small networks are not such balanced as expected from the consideration of triads only, they display significantly small degrees of unbalance (see the values of $\mathcal{U}$ in Fig. 2) in good agreement with Heider balance theory. Notice, that in many cases just the rewiring of a single link will produce a totally balanced network, e.g., rewiring the negative link between GB and Russia in the 1904 network to connect any of these two countries with any of AH, Ge or It, produces $\mathcal{U} = 0.0\%$.

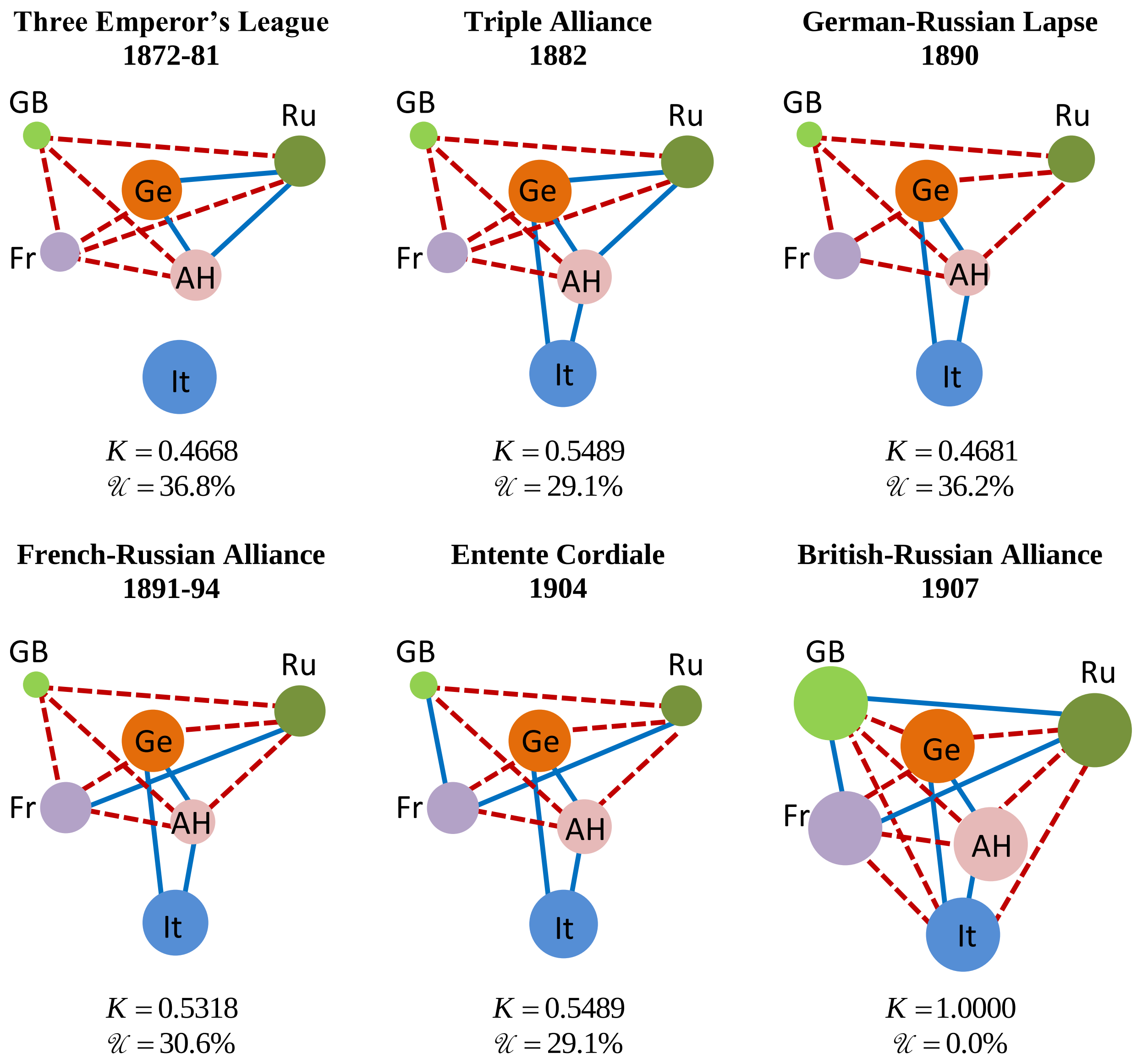

**Fig. 2.** Evolution of the global balance among the six major players of the World War I at different time periods. Solid blue lines account for alliances and broken red lines represent enmities. The degree of balance of every country is proportional to the radii of the circles. GB: Great Britain; Ru:

Russia; Ge: Germany; Fr: France; AH: Austro-Hungarian Empire; It: Italy.

The signed network of the Gahuku-Gama subtribe system of the Eastern Central Highlands of New Guinea describes a series of alliances and oppositions among the Gahuku-Gama subtribes,[27,28] which are distributed in a particular area and engage in prolonged warfare. In fact, "*Warfare…is that activity which characterizes the tribes of the Gahuku-Gama as a whole and which differentiates them from groups in other socio-geographic regions*".[27] The consideration of triangles only to account for the balance of the Gahuku-Gama network does not reveal all the interesting features of this alliance/conflict system. For instance, the index $K_3 = 0.735$ indicates that the network is in a *close-to-balanced* state. This is exactly what is revealed by the consideration of the node contribution to balance, which indicates that 5 (Gaveve, Ove, Alikadzuha, Nagamo and Ukudzuha) out of 16 subtribes are perfectly triangle-balanced. However, the spectral balance index is: $K = 0.335$, which points out to a state not so close to a balance, i.e., $\mathscr{U} = 49.8\%$. Notice that with the given number of nodes, positive and negative edges many networks can be constructed for which $\mathscr{U} = 0.0\%$. The lack of balance in this network is clearly extended beyond the triads. For instance the tribe of Nagam which is triad-balanced has a relatively poor node balance index of $K_i = 0.424$ due to the fact that it participates in several unbalanced squares and pentagons (see SI).

We have then considered the node balance index of each of the tribes in the Gahuku-Gama system. The tribes with largest balance are: Alikadzuha, Ove, Gaveve, Ukudzuha and Kotuni. All these tribes are geographically located in the Northeastern part of the Gahuku-Gama region. In contrast, the tribes displaying more unbalance are Uheto, Seuve, Notohana, Gehamo and Kohika, all of which (except Seuve) are located in the Western part of the Gahuku-Gama region. In the Fig. 3 it can be seen that there is a clear divisor line between the Southwestern and Northeastern parts of the region in terms of the local balance of the respective tribes. In this particular scenario it looks like the balance/unbalance is very much controlled by geographical constraints. The most unbalanced subtribe is that of Uheto, which is the one most to the Southwest of the region. By removing all links (positive and negative) which are incident to this subtribe, the global balance of the network increases from $K = 0.335$ ($\mathscr{U} \approx 49.8\%$) to $K = 0.532$ ($\mathscr{U} \approx 30.5\%$). This is equivalent to 'isolate' this subtribe from any other with which it has alliances or enmities to increase significantly the global balance of the system. All in all, this network, used as a classical example of balance in social relations according to Heider theory, is not so balanced as expected from the consideration of triads only. Although we can consider the previous small networks as relatively balanced, this one can only be considered as a moderately balanced network.

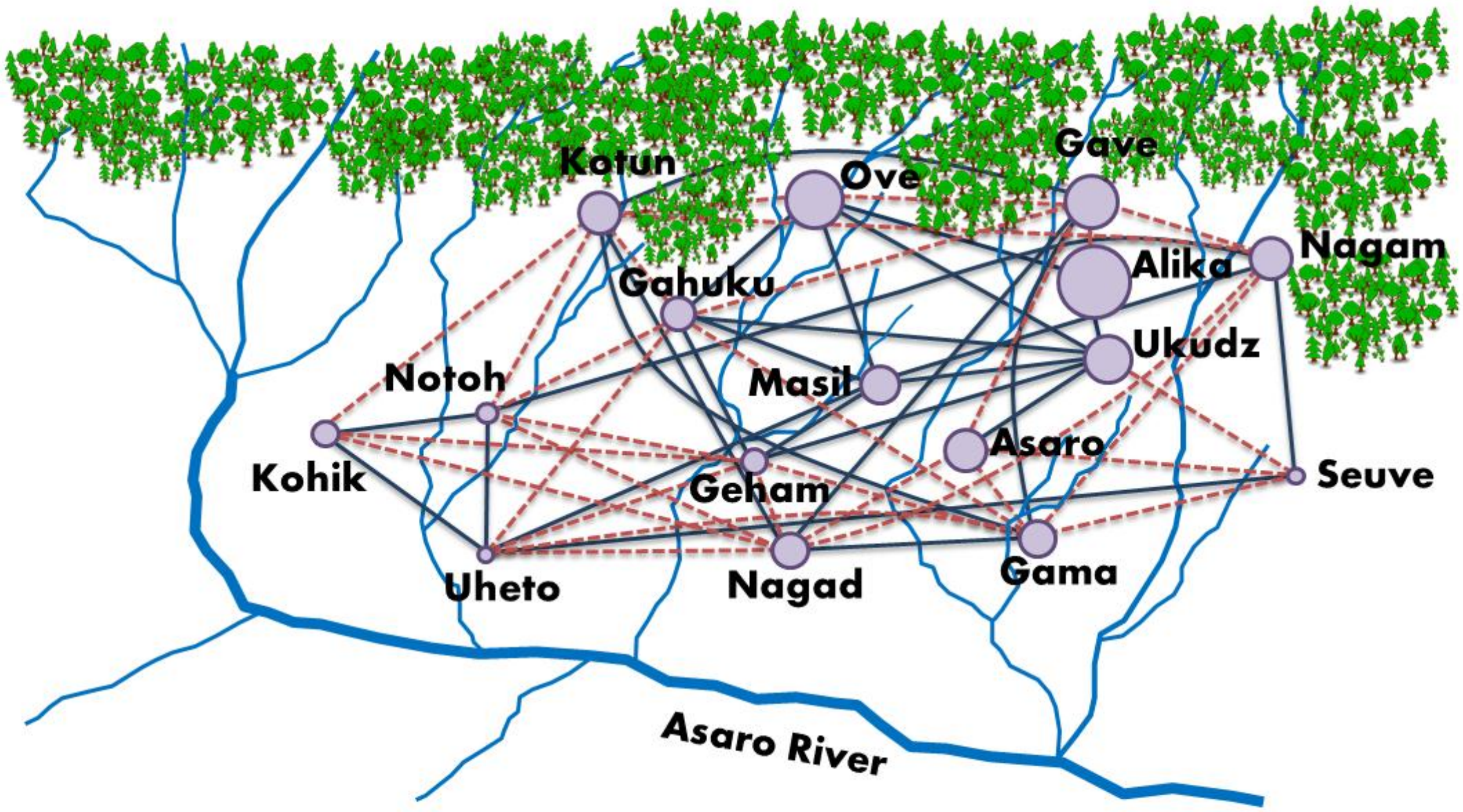

**Fig. 3.** Lack of global and local balance among the subtribes in the highlands of New Guinea. Subtribes are represented by circles with radii proportional to their degree of balance and located on an artistic representation of New Guinea highlands according to Read.[28] Continuous dark blue lines are for alliance ("rova") relations, and red discontinuous for antagonistic ("hina") relations.

**Balance in large online social networks.** The results obtained by considering $K_3$ and $K$ for three large online social networks: Epinions,[29] Slashdot (Zoo feature),[9,30] and Wikipedia,[31] are given in Table 1. Facchetti et al.[10] considered undirected versions of these networks and observed that Epinions, Slashdot and WikiElections have high percentages of balanced nodes (see Table 1). On the basis of these results, they concluded that these online social networks are highly balanced. The index $K_3$, which considers triads only, exactly reproduces this trend of high balance for the undirected versions of these networks. This demonstrates that the method used by Facchetti et al.[10] gives significantly more weight to the contributions of triads to the degree of balance in these networks. In contrast, our current results obtained by using the walk-balance index $K$ shows that the undirected versions of these three online networks are highly unbalanced, with percentages of balance not far from 0% (see Table1). More interestingly, the analysis of the directed networks shows that with the exception of Epinions, the other two online social networks are very much unbalanced. Notice that, according to the unbalance index $\mathscr{U}$, Slashdot and WikiElections have 87.1% and 99.99% of unbalance in their structures. As before, it is worth mentioning that there are many balanced networks that can be constructed by rewiring the positive and negative links of these networks. The only think that should be done is to split the nodes of these networks into two sets. Then, use the negative links only to connect nodes in the two different sets and positive links to connect nodes inside the same set. The resulting networks will display perfect balance by definition.[20]

**Table 1. Balance in signed online social networks**

| | | Undirected | | | Directed | | |
|---|---|---|---|---|---|---|---|
| Network | % Bal.[a] | $K_3$ | $K$ | $\mathscr{U}$ (%) | $K_3$ | $K$ | $\mathscr{U}$ (%) |
| Epinions | 83.7 | 0.808 | $1.88\cdot10^{-15}$ | 100 | 0.759 | 0.761 | 13.6 |
| Slashdot | 68.3 | 0.772 | $2.63\cdot10^{-7}$ | 100 | 0.880 | 0.069 | 87.1 |
| WikiElections | 52.9 | 0.595 | $3.29\cdot10^{-12}$ | 100 | 0.511 | $2.22\cdot10^{-5}$ | 99.99 |

[a]Percentage of balanced nodes reported by Faccetti et al.[10]

The three online social networks are, however, more balanced than expected from a random allocation of the signs to the edges. We have randomly reshuffled the signs of the edges in these networks keeping the exact proportion of positive to negative links. The randomly reshuffled networks display significantly less balance than the real ones: $K \sim 10^{-17}$ (Slashdot), $K \sim 10^{-18}$ (Epinions) and $K \sim 10^{-9}$ (WikiElections). This result indicates that the real-world networks are more balanced than expected from a totally random allocation of the friendship/enmity relations among the people. Taking the two results together we should conclude that the online social networks of Slashdot and WikiElections are very far from the ideal balance predicted by Heider theory, although they are more balanced than expected from a random allocation of the edge signs.

Where do these high levels of structural unbalance come from? Leskovec et al.[8] have found that triads of the form "*the enemy of my enemy is my friend*" are significantly underrepresented in these three online social networks. The triads with only one negative link or with all three negative links have been found to be overrepresented in the three online networks.[8] Because the triad with only one negative link is unbalanced it is plausible to ask whether the high unbalance is coming mainly from the all-negative triads. To respond that question we constructed the sub-networks of the online networks in which only the negative links are considered. Here we will describe the results only for the directed networks (see SI for the results on undirected versions of the networks). The three negative sub-networks display very poor degrees of balance ($\mathscr{U} = 100\%$ for Epinions, $\mathscr{U} = 91.4\%$ for Slashdot and $\mathscr{U} = 84.3\%$ for WikiElections), which are not significantly different from the ones obtained by random reshuffling of the networks and further extraction of the negative sub-networks ($\mathscr{U} = 100\%$ for Epinions, $\mathscr{U} = 100\%$ for Slashdot and $\mathscr{U} = 97.9\%$ for WikiElections). These results support the idea that a great deal of the unbalance in these online social networks comes from the totally negative cycles in the networks, supporting the previous findings of overrepresentation of all-negative triads in these networks.[8] However, not all of the unbalance comes from negative triads as should be expected from Davis's weaker notion of balance. If we compare the previous results with those obtained by using the index $K_3$, which accounts only for triads, we see a large contrast. In this case the three negative sub-networks display high degree of balance, which are larger than the ones obtained for the randomly reshuffled ones (in parenthesis): $K_3 = 0.265(0.045)$ for Epinions, $K_3 = 0.719(0.110)$ for Slashdot and $K_3 = 0.183(0.140)$ for WikiElections. Thus, the existence of many other negative cycles is responsible for that global lack of balance in these networks. Finding such individual negative fragments is a giant computational task due to the size of the networks and the typical combinatorial explosion of signed directed fragments in networks. However, our results are conclusive in determining that these social networks are not as balanced as expected from Heider balance theory. These levels of unbalance are not incompatible with Davis's model of weak balance if the model is modified to consider other types of negative fragments apart from the all-negative triads.

**Tuning balance in social networks.** A potential advantage of the consideration of the walk balance index as an equilibrium constant is that we can study the effects of the inverse temperature $\beta$ over the index. This represents a way to tune the degree of balance of a network without changing its topology. The inverse temperature plays the role here of an overall importance given to the opinions in a social network, i.e., high importance corresponds to $\beta \to \infty$ ($T \to 0$), while low importance implies $\beta \to 0$ ($T \to \infty$).

The plots of the equilibrium constant $K$ vs. $\beta$ for Slashdot and WikiElection networks follow an exponential decay. This plot corresponds to the network analogue of the van't Hoff plot[32] and the exponential dependence of $K$ with the $\beta$ for these two networks can be described by the network analogue of the van't Hoff equation $d \ln K / d\beta = -\frac{\Delta H^\circ}{R}$,[32] where $\Delta H^\circ$ is the standard enthalpy of the network transformation represented in Fig. 1. If we assume that $\Delta H^\circ$ is independent of $T$, the integration of the van't Hoff equation results in the well-known linear form:[33]

$$\ln K = -\frac{\Delta H^\circ}{R}\beta + \frac{\Delta S^\circ}{R}, \tag{6}$$

where $\Delta S^\circ$ is the standard entropy of the network transformation. A plot of $\ln K$ vs. $\beta$ gives a straight line with the intercept $\Delta S^\circ / R$ and the slope $-\Delta H^\circ / R$. The negative slopes obtained for both Slashdot and WikiElection networks indicate that the transformation from a totally balanced network to an unbalanced one is endothermic, i.e., the system absorbs energy from its surroundings. The slope for Slashdot is $-2.67$ and that for WikiElections is $-10.80$, indicating that the second is a significantly more 'endothermic' process than the first. In other words, obtaining the unbalance in the WikiElections network costs more 'energy' to the system than that for Slashdot. This result perfectly fits with the fact that in the Wikipedia network the edge signs are more public than in Slashdot. Thus, it is plausible that users, who can see the votes of others, have more tendency to conform to already positive voting outcomes as has been clearly remarked by Leskovec et al.[8] This inertia to vote negatively is represented in our model by a higher 'endothermicity' of the process of converting positive to negative links in the equilibrium depicted in Fig.1. Mathematically, the behavior of these two networks can be explained by the fact that the spectral gap of both $\Sigma$ and $|\Sigma|$ is relatively large, i.e., $\lambda_1(\Sigma) >> \lambda_2(\Sigma)$ and $\lambda_1(|\Sigma|) >> \lambda_2(|\Sigma|)$. Then, the equilibrium constant can be approximated by

$$K \cong \frac{\exp(\beta\lambda_1(\Sigma))}{\exp(\beta\lambda_1(|\Sigma|))} = \exp\left(-\beta\left(\lambda_1(|\Sigma|) - \lambda_1(\Sigma)\right)\right), \tag{7}$$

which displays the perfect exponential decay observed in the Fig. 4, i.e., the plots of $\ln K$ versus $\beta$ for these networks are perfect straight lines with correlation coefficients larger than 0.999.

The Epinions network displays a completely nonlinear behavior in its van't Hoff plot, which clearly points to a nonmonotonic change of the balance with the temperature. As can be seen in the Fig. 4 there is a local minimum at $\beta = 0.09$ ($K = 0.6058$) and then the absolute maximum is obtained at $\beta = 0.62$ ($K = 0.7996$) after which the balance decays exponentially. Before explaining the causes for this nonlinear behavior of the van't Hoff plot let us remark what it means in terms of network balance. It is usually assumed that balance in signed (social) networks depends uniquely on the sign pattern and the topological arrangement of the nodes and links in the network. Here we observe for the first time that the 'environmental' conditions in which these networks are embedded can change the balance in a nonmonotonic way. Suppose that every link in this network receives the typical weight of one. Then, the global balance of the network is $K = 0.761$. This balance can be increased if every link in the network receives a weight of $\beta = 0.62$, which could means, for instance, that we decrease the 'importance' of every opinion represented by a link in the network. However, further decreasing this weight will reduce the balance down to $K = 0.6058$ when $\beta = 0.09$. More importantly, increasing the weight we give to the opinions in the network beyond the typical value of one does not improve the balance but decreases it down to an asymptotic value of zero for $\beta \to \infty$.

The nonlinearity of van't Hoff plots is well documented for physical systems. It is a consequence of the lack of independence of $\Delta H^{\circ}$ with $T$. In this case the integration of the van't Hoff equation gives rise to a meromorphic function in terms of the temperature:[32]

$$\ln K = -\frac{\Delta H^{\circ}}{R}\beta + a\ln\beta^{-1} + b\beta^{-1} + c\beta^{-2} + \cdots + C\,, \tag{8}$$

where $C$ is a constant of integration. We have fitted the van't Hoff curve for Epinions using such expression (see SI) and have obtained $\frac{\Delta H^{\circ}}{R} = -0.4716$, which is significantly smaller than the values obtained for Slashdot and WikiElections. That is, Epinions needs to take significantly less 'energy' from the environment in order to reach the level of balance observed in the network.

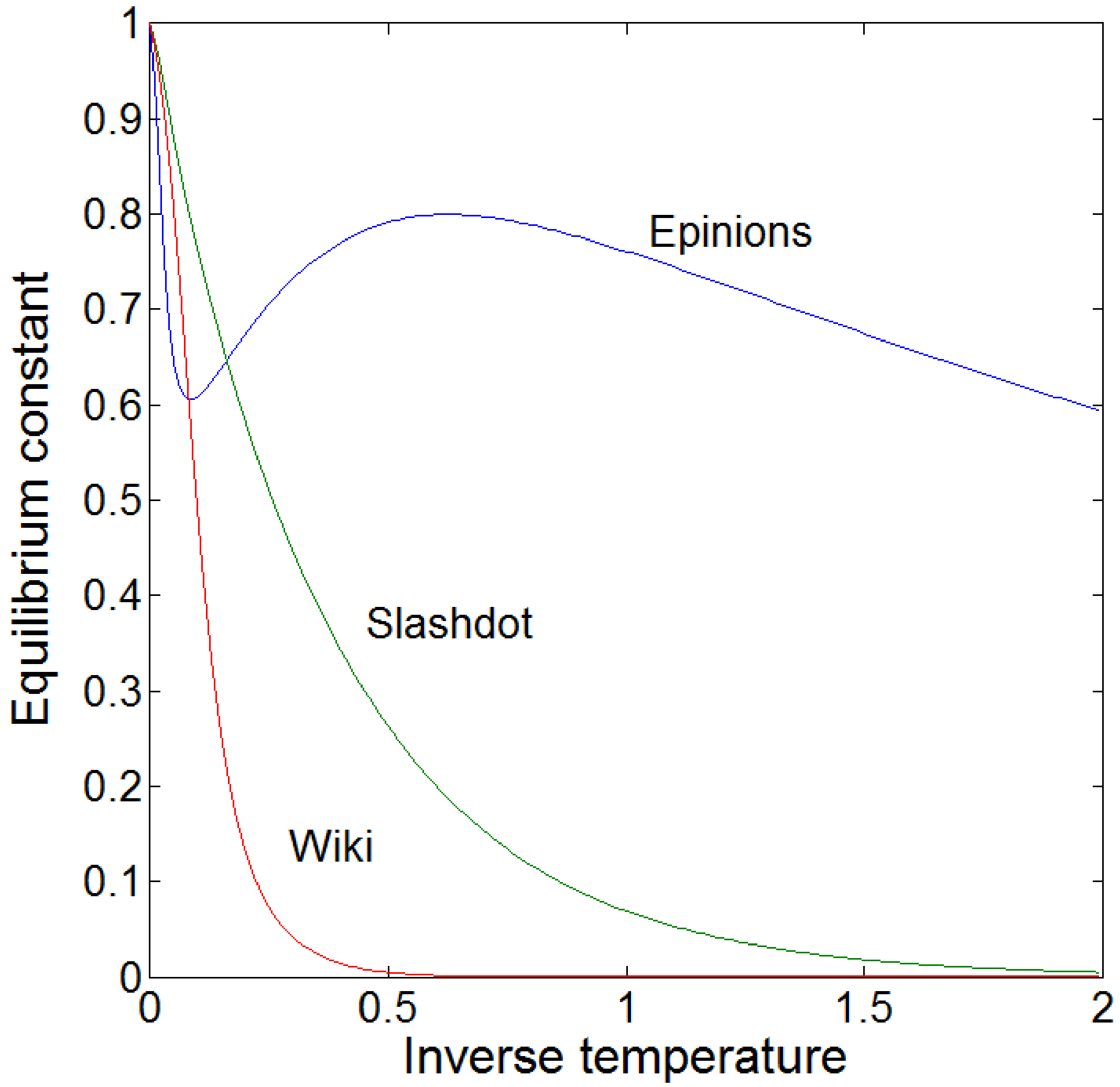

**Fig. 4.** Demonstration of the fact that the balance of a network can be dramatically changed by tuning the weights of its links. The plot represents the analogue of the van't Hoff plot for networks, where the equilibrium constant (balance index) is plotted against the inverse temperature (weight of the links).

## Discussion

We have developed a method that quantifies the global and local balance of a signed (social) network by accounting for the contribution of all cycles in the network, but giving more weights to the shorter than to the longer ones. This last requirement is based on empirical observations in social sciences. The degree of balance of a network is then obtained from the calculation of the spectra of their adjacency matrix, so that no ad hoc heuristic is needed. The walk-balance index can be understood as an equilibrium constant for a hypothetical dynamics in which some of the edges of an all-positive network becomes negative. This formulation allows the introduction of an important parameter, the temperature, which modulates the relative importance given to the opinions in a social network.

We have observed that real-world social networks from very different scenarios are in general not as balanced as expected from Heider theory. These results contrasts significantly with previous findings that social networks are in general extremely balanced.[10] The main differences could be due to the fact that here we consider balance from a wider structural perspective in which all potential cycles make a contribution to the balance/unbalance of a network, but in which we also account for the empirical observation that shorter cycles have a larger influence over balance than longer ones. Small cycles different from triads, like signed squares, are also important for describing the balance in networks and consequently for elaborating theories that explain observed lack of balance in certain social networks. Another important idea put forward in this work is that balance can be modified by changing the weights of the links in a network by using the physical metaphor of a network temperature. This allows the modification of the balance state of a network without changing its topology at all. Consequently, diminishing the possibilities of conflicts in such kind of friend/enmity networks is possible by tuning the importance given to the general opinions expressed in those networks. Thus, taking together all of its aspects, our method for describing balance offers deeper understanding of the structural and dynamical nature of balance in signed social networks.

## Materials and Methods

**Datasets.** The signed social networks analyzed in this work are: (i) Gama, a set of political alliances and oppositions among the Gahuku-Gama subtribes in the highland New Guinea,[27,28] (ii) WWI, networks of relations among the major players in the First World War at different times,[7] (iii) Epinions, trust/distrust network among users of the product review site Epinions,[29] (iv) WikiElections, network representing the votes for the election of administrators in Wikipedia,[31] (v) Slashdot, network of friend/foe in the technological news site Slashdot.[9,30] The network (i) was downloaded from UCINET IV Datasets at http://vlado.fmf.uni-lj.si/pub/networks/data/ucinet/ucidata.htm, (ii) was built from the information provided in the ref. (7), and (iii)-(v) were downloaded from the Stanford Network Analysis Platform (http://snap.stanford.edu/). The number of nodes and signed links of the three online social networks are given in Table 2.

**Table 2. Number of nodes and signed links in three online social networks**

| Network | $n$ | $m^{+}$ | $m^{-}$ |
|---|---|---|---|
| Epinions | 131,828 | 717,667 | 123,705 |
| Slashdot | 82,144 | 425,072 | 124,130 |
| WikiElections | 8,297 | 81,664 | 21,927 |

**Computational approaches**. All calculations were performed using Matlab. While the calculation of $K$ does not involve any difficulties when the networks are small, large-scale networks require advanced computational techniques. For the three online social networks studied here we have used the Implicitly Restarted Arnoldi (IRA) method.[33,34] This algorithm can be used to compute a user-specified number k of selected eigenvalues $\lambda_1, \lambda_2, \ldots, \lambda_k$ of largest magnitude of the input matrices $A(\Sigma)$ and $A(|\Sigma|)$. We then approximate $tr(\exp(M)) \cong \sum_{j=1}^{k} \exp(\lambda_j)$, where $M$ is either $A(\Sigma)$ or $A(|\Sigma|)$. When the matrices $A(\Sigma)$ and $A(|\Sigma|)$ are nonsymmetric, like in the case of directed networks, there are some eigenvalues which are non-real. Hence, it is possible in principle that the approximation $tr(\exp(M)) \cong \sum_{j=1}^{k} \exp(\lambda_j)$, will have a nonzero imaginary part. This problem can be easily avoided by observing that the approximation will be real provided that we include the conjugate $\bar{\lambda}_j$ of every complex $\lambda_j$ among the $k$ eigenvalues used in the approximation, since in this way the imaginary parts of $\exp(\beta\lambda_j)$ and $\exp(\beta\bar{\lambda}_j)$ will cancel each other out. Moreover, in our calculations we found that the few eigenvalues of largest magnitude tend to be real, so that the computed approximations are either real or have small imaginary part, which can be simply ignored. In practice, we found that very small values of $k$ give excellent approximations, owing to the fact that the eigenvalues of largest magnitude have positive real part and are well-separated from the rest of the spectrum. Experimenting with different values of $k$ shows that increasing $k$ above a small, fixed value does not appreciably change the value of the traces, in relative terms. Values of $k$ between 6 and 10 yield tiny relative errors, but in some cases even $k = 1$ results in an acceptable approximation. In summary, we found that the IRA method provides a very effective approach for approximating the trace of the exponential of large adjacency matrices of both signed and unsigned networks with the complexity being approximately $O(n)$ if k is small and fixed.

**ACKNOWLEDGEMENTS.** EE acknowledge the Royal Society for a Wolfson Research Merit Award. MB's work was supported by National Science Foundation grant DMS1115692.

# Supplementary Information

## 1. Analytic Results

We prove a result about the existence of graphs for which the global balance tends to zero as the number of nodes tends to infinite. Here, $C_n$ and $K_n$ stand for the cycle and complete graphs, respectively. The cycle is the graph in which all the nodes are connected to two other nodes. The complete graph is the graph in which every pair of nodes is connected by an edge.

**Theorem 1**. Let $G_n$ be the graph whose adjacency matrix is given by:

$$A(G_n) = 2A(C_n) - A(K_n). \quad \text{(S1)}$$

Then, $K(G_n) \to 0$ as $n \to \infty$.

***Proof***. First we start by proving that the matrices $A(C_n)$ and $A(K_n)$ commute. Let $E = \mathbf{1} \cdot \mathbf{1}^T$ where $\mathbf{1}$ is an all-ones vector. Obviously, $A(K_n) = E - I$, where $I$ is the corresponding identity matrix. Then, $A(K_n) \cdot A(C_n) = E \cdot A(C_n) - A(C_n)$ and $A(C_n) \cdot A(K_n) = A(C_n) \cdot E - A(C_n)$. The two matrices commute if $A(K_n) \cdot A(C_n) = A(C_n) \cdot A(K_n)$, which implies that $E \cdot A(C_n) = A(C_n) \cdot E$. It can be easily checked that

$$E \cdot A(C_n) = [k_1\mathbf{1} \quad k_2\mathbf{1} \quad \cdots \quad k_n\mathbf{1}], \quad \text{(S2)}$$

and

$$A(C_n) \cdot E = \begin{bmatrix} k_1\mathbf{1}^T \\ k_2\mathbf{1}^T \\ \vdots \\ k_n\mathbf{1}^T \end{bmatrix}, \quad \text{(S3)}$$

where $k_i$ is the degree of the node $i$. Then, if the graph is regular, $k_1 = k_2 = \cdots = k_k = r$ and $A(C_n) \cdot E = E \cdot A(C_n) = r[\mathbf{1} \quad \cdots \quad \mathbf{1}]$, which proves that the adjacency matrix of a complete graph and that of any regular graph commute. Because the cycle is a regular graph, the first part of the proof is complete.

Because of the commutativity between the adjacency matrices of the cycle and complete graph we can start by writing the ratio

$$\frac{Z(G_n)}{Z(|G_n|)} = \frac{tr[\exp(2A(C_n)) \cdot \exp(-A(K_n))]}{tr[\exp(2A(C_n)) \cdot \exp(A(K_n))]} . \tag{S4}$$

Using the eigenvalues and eigenvectors of the adjacency matrices of cycles and complete graphs we have

$$(\exp(A(C_n)))_{pp} = \frac{1}{n}\sum_{j=0}^{n/2} e^{2\cos\left(\frac{2\pi j}{n}\right)} , \tag{S5}$$

$$(\exp(A(C_n)))_{pq} = \frac{1}{n}\sum_{j=0}^{n/2} e^{2\cos\left(\frac{2\pi j}{n}\right)} \cos\left(\frac{2\pi j(p-q)}{n}\right), \tag{S6}$$

$$(\exp(A(K_n)))_{pp} = \frac{e^{n-1}}{n} + \frac{n-1}{ne} , \tag{S7}$$

$$(\exp(A(K_n)))_{pq} = \frac{e^{n-1}-1}{ne} , \tag{S8}$$

$$(\exp(-A(K_n)))_{pp} = \frac{1}{ne^{n-1}} + \frac{(n-1)e}{n} , \tag{S9}$$

$$(\exp(-A(K_n)))_{pq} = \frac{1}{ne^{n-1}} - \frac{e}{n} . \tag{S10}$$

For $j = 1,2,\cdots,n$ the angles $j\pi/(n+1)$ uniformly cover the interval $[0,\pi]$, thus enabling the usage of the following integral approximation:

$$(\exp(A(C_n)))_{pp} \approx \frac{1}{\pi}\int_0^{\pi} e^{2\cos\theta} d\theta = I_0(2), \tag{S11}$$

$$(\exp(A(C_n)))_{pq} \approx \frac{1}{\pi}\int_0^{\pi} e^{2\cos\theta} \cos(\theta(p-q)) = I_{d(p,q)}(2), \tag{S12}$$

where $I_\alpha(x)$ is the Bessel function of the first kind and $\alpha = d(p,q)$ is the shortest path distance between the nodes $p$ and $q$ in the network.

Then, using the fact that

$$\sum_{j=1}^{\infty} I_j(2) = \frac{1}{2}\left(e^2 - I_0(2)\right), \tag{S13}$$

we finally obtain

$$\lim_{n\to\infty} K(G_n) = \lim_{n\to\infty} \frac{\left[\frac{1}{ne^{n-1}} + \frac{(n-1)e}{n}\right](nI_0(2)) + \left(\frac{1}{ne^{n-1}} - \frac{e}{n}\right)\sum_{j=1} I_j(2)}{\left[\frac{e^{n-1}}{n} + \frac{(n-1)}{ne}\right](nI_0(2)) + \left(\frac{e^{n-1}-1}{ne}\right)\sum_{j=1} I_j(2)}$$

(S14)

$$= \lim_{n\to\infty} \frac{\left[\frac{1}{e^{n-1}} + (n-1)e\right](I_0(2)) + \left(\frac{1}{ne^{n-1}} - \frac{e}{n}\right)\left(\frac{e^2 - I_0(2)}{2}\right)}{\left[e^{n-1} + \frac{(n-1)}{e}\right](I_0(2)) + \left(\frac{e^{n-1}-1}{ne}\right)\left(\frac{e^2 - I_0(2)}{2}\right)} = 0.$$

## 2. Numerical Results

### 2.1 All-negative undirected sub-networks

**Table S1.** Balance indices in the all-negative sub-networks of the undirected versions of the three online social networks studied.

| Network | $K$ | $K$ (rnd) | $K_3$ | $K_3$ (rnd) | $\bar{C}$ [a] | $\bar{C}$ [a] (rnd) |
|---|---|---|---|---|---|---|
| Epinions | $4.10 \cdot 10^{-11}$ | $\sim 10^{-4}$ | 0.652 | 0.681 | 0.012 | 0.022 |
| Slashdot | $1.38 \cdot 10^{-6}$ | 0.025 | 0.758 | 0.851 | 0.005 | 0.010 |
| WikiElections | $3.95 \cdot 10^{-5}$ | $\sim 10^{-6}$ | 0.569 | 0.890 | 0.028 | 0.031 |

[a]Average Watts-Strogatz clustering coefficient reported by Leskovec et al.[8]

### 2.2 Fit of van't Hoff Equations for the Online Social Networks

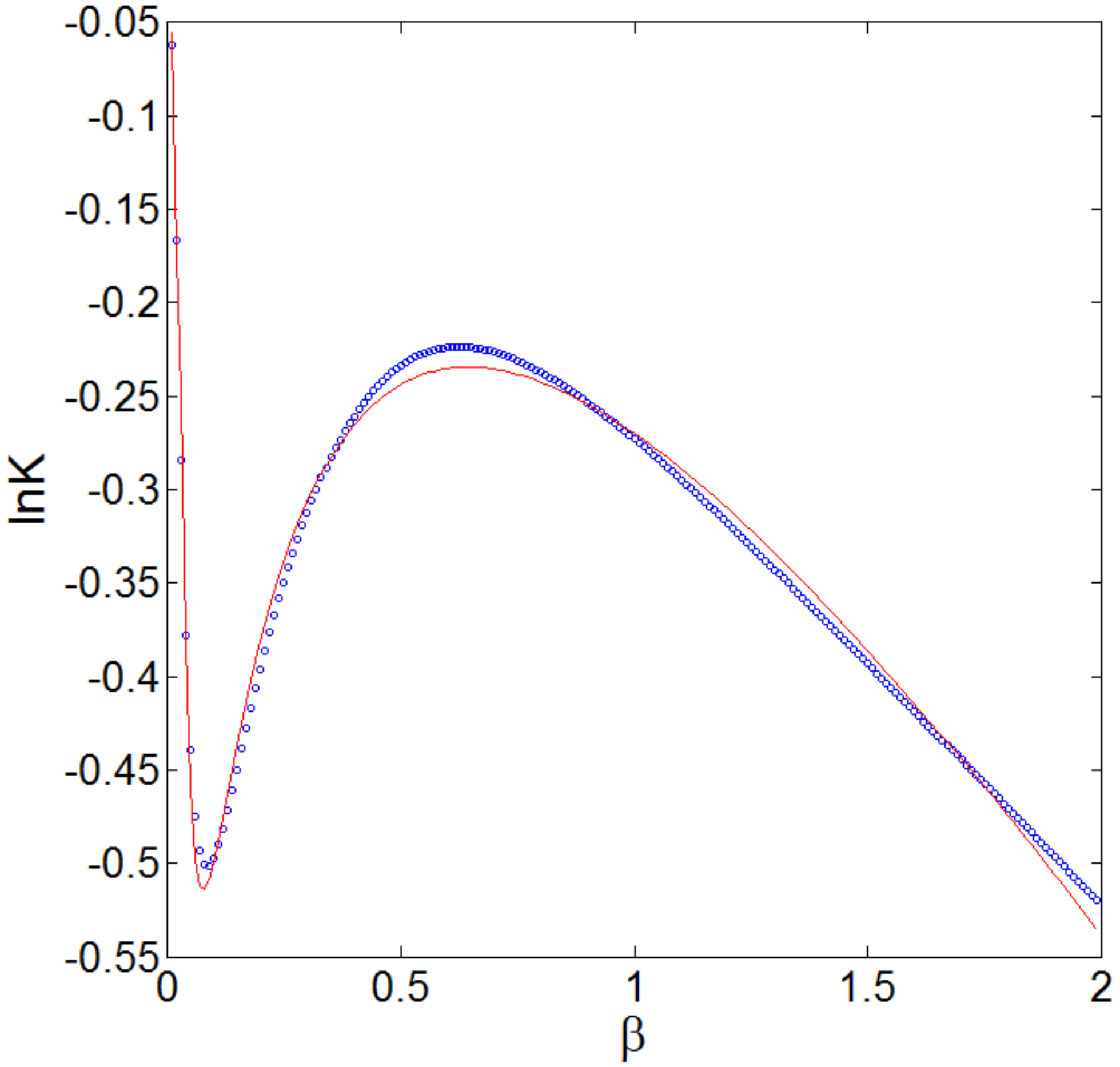

**Fig. S1.** Nonlinear change of the balance ($\ln K$) with the weight of the links (inverse temperature, $\beta$) in the online social network Epinions. The circles represent the values from the simulation and the solid line represents the fit using

$\ln K = -\frac{\Delta H^{\circ}}{R}\beta + a\ln\beta^{-1} + b\beta^{-1} + c\beta^{-2} + \cdots + e\beta^{-5} + C$, where $\frac{\Delta H^{\circ}}{R} = -0.4716$, $a = -0.279$, $b = -0.02308$, $c = 0.002422$, $d = -4.625\cdot 10^{-5}$, $e = 2.535\cdot 10^{-7}$, and $C = 0.2223$. The squared correlation coefficient and root of the mean standard error are, respectively: $R^2 = 0.9922$ and $RMSE = 0.008625$.

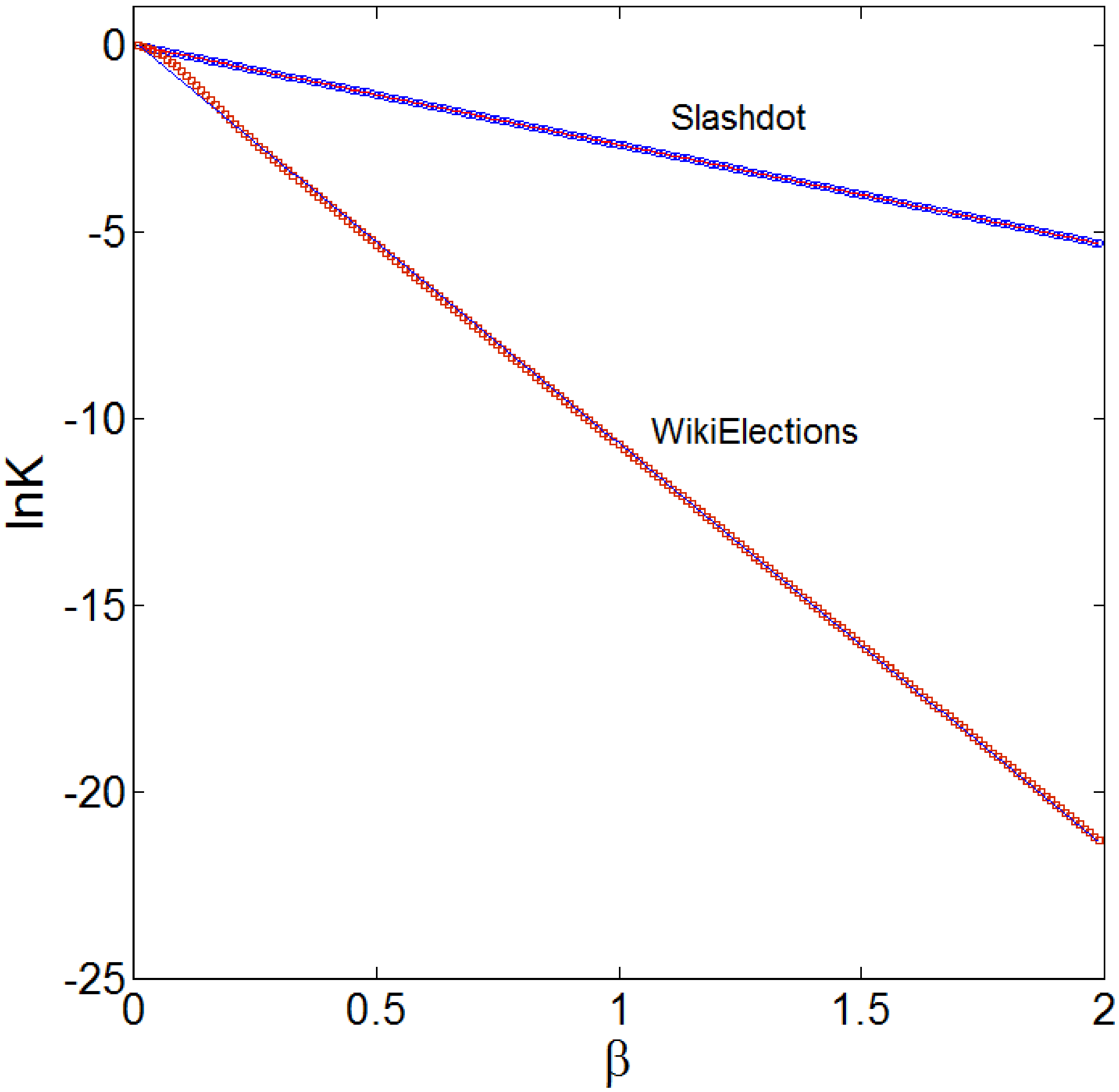

**Fig. S2.** Linear change of the balance ($\ln K$) with the weight of the links ($\beta$) in the online social networks WikiElections and Slashdot. The circles and squares represent the values from the simulation and the solid lines represent the fits using the $\ln K = a\beta + b$, where $a = -\frac{\Delta H}{R}$ and with the parameters given in Table S2.

**Table S2.** Fitting parameters for the van't Hoff plots of the online social networks of Slashdot and Wikielections.

| Network | $a$ | $b$ | $R^2$ | $RMSE$ |
|---|---|---|---|---|
| Slashdot | $-2.676$ | 0.001023 | 1.0000 | 0.001611 |
| WikiElections | $-10.8$ | 0.1161 | 0.9999 | 0.06957 |

$R^2$ is the squared correlation coefficient and $RMSE$ is the root of the mean standard error.